\documentclass[a4paper,fleqn,usenatbib]{mnras}

\usepackage{txfonts}

\usepackage[T1]{fontenc}
\usepackage{ae,aecompl}

\usepackage{psfig}   
\usepackage{graphicx}
\usepackage{amssymb}
\usepackage{subfigure}

\title[Core mass segregation]{On the mass segregation of cores and stars}

\author[H. L. Alcock and R. J. Parker]{Hayley L. Alcock and Richard  J. Parker\thanks{E-mail: R.Parker@sheffield.ac.uk}\thanks{Royal Society Dorothy Hodgkin Fellow} \vspace*{0.1cm}\\
Department of Physics and Astronomy, The University of Sheffield, Hicks Building, Hounsfield Road, Sheffield, S3 7RH, UK}

\begin{document}

\date{}
                             
\pagerange{\pageref{firstpage}--\pageref{lastpage}} \pubyear{2019}

\maketitle

\label{firstpage}

\begin{abstract}
Observations of pre-/proto-stellar cores in young star-forming regions show them to be mass segregated, i.e.\,\,the most massive cores are centrally concentrated, whereas pre-main sequence stars in the same star-forming regions (and older regions) are not. We test whether this apparent contradiction can be explained by the massive cores fragmenting into stars of much lower mass, thereby washing out any signature of mass segregation in pre-main sequence stars. Whilst our fragmentation model can reproduce the stellar initial mass function, we find that the resultant distribution of pre-main sequence stars is mass segregated to an even higher degree than that of the cores, because massive cores still produce massive stars if the number of fragments is reasonably low (between one and five). We therefore suggest that the reason cores are observed to be mass segregated and stars are not is likely due to dynamical evolution of the stars, which can move significant distances in star-forming regions after their formation.  
\end{abstract}


\section{Introduction}

Understanding the process through which stars form from the collapse and fragmentation of giant molecular clouds is one of the great unsolved puzzles of modern astrophysics. Whilst it is an interesting topic of study in its own right, the problem is fundamentally important for understanding the early phases of planet formation \citep{Bonnell01b,Haisch01,Adams06}, including our own Solar system \citep{Adams10}, as well as for understanding the formation and evolution of galaxies \citep{Keres09,Krumholz18}.

Most stars appear to form in groups where the stellar density exceeds that in the Galactic disc by up to several orders of magnitude \citep{Lada03,Bressert10,Kruijssen12b}. These groups appear to form with structural \citep[e.g.][]{Larson95,Cartwright04,Gouliermis14} and kinematic substructure \citep[e.g.][]{Alfaro16,Wright16}. This substructure governs the long-term evolution of these star-forming regions \citep{Parker14b,Sills18}, but it is unclear how or why stars inherit these properties from their parent molecular clouds.

Advances in submillimetre observations in the past decade have shown that stars form from overdensities in GMCs, often referred to as pre-stellar or protostellar cores\footnote{In this paper, we will not distinguish between pre-stellar, protostellar, or starless cores.}, depending on their evolutionary stage \citep{Andre00,Alves07,Andre10,Arzoumanian11}. Not all of these observed cores are gravitationally bound, so only a subset are likely to form stars \citep[e.g.][]{Caselli95}. Studies of the core mass function (CMF) show that it has a similar form to the stellar initial mass function (IMF), but offset by a factor of three towards higher masses \citep{Alves07}. This has led to considerable debate about whether the intrinsic properties of stars are set by the cores, or whether the cores subsequently fragment into stars that have very little memory of the physics within the cores \citep[e.g.][and references therein]{Padoan02,Hennebelle08b,Hopkins12,Holman13,Offner14}.

Despite their relatively short lifetimes, pre/proto-stellar cores are now observed in significant numbers in nearby star-forming regions such that -- in addition to measuring their mass distributions -- we can also quantify the spatial distributions of these objects, and compare these distributions to those of young (pre-main sequence) stars.

Several authors \citep{Kirk16b,Alfaro18,Parker18a,Dib18b,Plunkett18} have shown that the spatial distributions of cores in various star-forming regions tend to be mass segregated, where the most massive cores are more centrally concentrated than would be expected from a random distribution. Mass segregation is one of the predictions of the competitive accretion theory of star formation \citep{Zinnecker82,Bonnell97,Bonnell98}, though it is usually assumed that this arises from accretion of gas onto Jeans-mass fragments in the central regions of a forming star cluster, rather than as a direct consequence of the spatial distribution of cores.

Interestingly, many star-forming regions do not exhibit mass segregation of pre-main sequence stars \citep{Parker11b,Parker12c,Parker17a,Gennaro17,Dib18}, though the Orion Nebula Cluster is a notable exception \citep{Hillenbrand98,Allison09a}.

Furthermore, \citet{Plunkett18} find that whilst the pre-stellar cores in Serpens South are mass segregated, the pre-main sequence stars are not. This raises the interesting question of why mass segregation appears in the earliest phases of star-formation, but appears to dissipate later on \citep[see also][]{Elmegreen14}. There are four possible explanations for this. First, observations of cores may be biased to observing mass segregation, whereas observations of stars are not. Second, the cores that are forming today could have a different spatial distribution to those that formed the pre-main sequence stars we observe today. Third, dynamical evolution of the pre-main sequence stars may have moved them away from their birth sites and erased any signature of mass segregation. Fourth, the fragmentation process within cores could produce stars randomly with no preferred spatial distribution. In this process, we might expect that massive cores would fragment into many lower-mass objects, thus erasing the mass segregation signal of the cores.

In this paper, we investigate the fourth scenario, i.e.\,\,that the fragmentation of the pre-/proto-stellar cores erases the primordial mass segregation of these objects. (We will also discuss the other scenarios in light of our results later in the paper.) The paper is organised as follows. In Section~\ref{methods} we describe the set-up and assumptions behind a simple toy-model to convert cores into stars through fragmentation. We present our results in Section~\ref{results}, we provide a discussion in Section~\ref{discuss} and we conclude in Section~\ref{conclude}.

\section{Methods}
\label{methods}

In this section we describe the set-up of our synthetic distributions of cores and how we allow those cores to fragment into pieces (which we assume are stars). We assume that each fragment will subsequently form a pre-main sequence star, and that all of the mass in the fragment ends up in the star.

\subsection{Spatial distribution of cores}

In order to mimic the filamentary nature of star formation, and the observed substructured distribution of cores, we distribute the cores in a fractal with a high degree of substructure. We use the box fractal method described in \citet{Goodwin04a}, \citet{Cartwright04}, \citet{Allison10} and \citet{Parker14b}, and for a full description of the method we refer the interested reader to those works. In brief, the fractal is constructed by placing `parent' particles at the centre of cubes, and the number of parent particles that produce children is set by the fractal dimension, $D$. For a low fractal dimension, the parents produce fewer children and so the resultant distribution has more substructure.

We adopt a fractal dimension $D = 1.6$ for our core distributions, which results in a highly substructured distribution. However, we have repeated the analysis with a centrally concentrated spherical distribution with a density profile $n \propto r^{-2.5}$ and our overall results are very similar. This is because our method for quantifying mass segregation does not assume, or depend on, anything about the underlying spatial distribution. The star-forming regions have radii of 1\,pc, but this does not affect how mass segregation is quantified using the dimensionless $\Lambda_{\rm MSR}$ ratio \citep{Allison09a}.

\subsection{Masses of cores and fragments}

Following the observations of \citet{Alves07}, we assume that the core mass function is the same as the stellar initial mass function, but shifted to higher masses by a factor of $\sim 3$. We therefore sample from the \citet{Maschberger13} Initial Mass Function, which has a probability distribution of the form
\begin{equation}
  p(m) \propto \left(\frac{m}{\mu}\right)^{-\alpha}\left(1 + \left(\frac{m}{\mu}\right)^{1 - \alpha}\right)^{-\beta},
  \label{imf}
\end{equation}
but we adopt $\mu = 0.6$\,M$_\odot$ as the average \emph{core} mass, and sample masses from this function in the range 0.3 -- 150\,M$_\odot$. As the shape of this distribution is assumed to be the same as the stellar initial mass function, we adopt $\alpha = 2.3$ \citep{Salpeter55} and $\beta = 1.4$.

We sample $N_{\rm cores} = 300$ from Equation~\ref{imf} and we then allow the cores to fragment according to the following prescriptions:\\

\noindent {\it Random fragmentation:} the cores are randomly fragmented into between one and five equal-mass pieces.\newline
{\it Low-mass fragmentation:} the low-mass ($< 1$M$_\odot$) cores fragment into five equal-mass pieces; the massive cores fragment into one piece.\newline
{\it High-mass fragmentation:} the high-mass ($\geq 1$M$_\odot$) cores fragment into five equal-mass pieces; the low-mass cores fragment into one piece.\\

Depending on the type of fragmentation, we end up with between 710 and 1100 fragment pieces.

\subsection{Core mass segregation}

In one set of simulations, the cores are randomly assigned masses drawn from Equation~\ref{imf}. In all other simulations, the cores are mass segregated so that the core closest to the centre of the star-forming region has the highest mass, and so on, such that the least massive core is the most distant object from the centre. \citep[Other methods to mass segregate star-forming regions are available, e.g.][ but as we do not require the cores to have velocities, our simple geomtrical construct is sufficient]{Subr08}.

\subsection{Quantifying mass segregation}

We quantify mass segregation using the \citet{Allison09a} $\Lambda_{\rm MSR}$ method. This has the advantage over other techniques for quantifying mass segregation that it does not make any assumptions about the geometry of the star-forming region, and is therefore ideal for quantifying mass segregation in highly substructured distributions, such as those in our simulations.

The $\Lambda_{\rm MSR}$ ratio is calculated by drawing a minimum spanning tree between all of the objects in a chosen subset, which has length $l_{\rm subset}$. The number of objects in a chosen subset is $N_{\rm MST}$, and $\Lambda_{\rm MSR}$ is defined as the average length of the minimum spanning tree of $N_{\rm MST}$ objects, $\langle l_{\rm average} \rangle$, divided by the length of the subset $l_{\rm subset}$, thus:
\begin{equation}
\Lambda_{\rm MSR} = {\frac{\langle l_{\rm average} \rangle}{l_{\rm subset}}} ^{+ {\sigma_{\rm 5/6}}/{l_{\rm subset}}}_{-{\sigma_{\rm 1/6}}/{l_{\rm subset}}}.
\end{equation}
We conservatively estimate the lower (upper) uncertainty on $\Lambda_{\rm MSR}$ as the MST length which lies 1/6 (5/6) of the way through an ordered list of all the random lengths (corresponding to a 66 per cent deviation from the median value, $\langle l_{\rm average} \rangle$).

Variations of $\Lambda_{\rm MSR}$ exist, including those that quantify the geometric mean, rather than the arithmetic mean \citep{Olczak11,Dib18}, but for the purposes of this work the original definition is sufficient.

\subsection{Spatial distribution of fragments}

Depending on the fragmentation (see above), a core will fragment into between one and five pieces. In most of the simulations, the fragments simply inherit the position of their parent core, with a small (up to $\sim 0.05$\,pc) random offset applied. In one set of simulations we apply a more significant random offset (up to 0.5\,pc) to mimic the effects of either dynamical ejection from the core \citep{Reipurth01,Goodwin05c} or migration in the star-forming region of the fragments (i.e.\,\,as they become pre-main sequence stars and move within the gravitational potential of the star-forming region).  \\

A summary of the different simulations is provided in Table~\ref{sim_summary}.

\begin{table}
  \caption[bf]{Summary of the different spatial distributions of cores and fragments. The first column indicates whether the cores are mass sgeregated. The second column indicates the type of fragmentation; either random, or where the low-mass cores fragment and high-mass ($>1$\,M$_\odot$) do not, or where the high-mass cores fragment and the low-mass cores do not. The third column indicates whether a spatial offset is applied to the fragments to mimic the effects of dynamical evolution. The fourth column indicates the figures in Section~\ref{results} that show each particular simulation.} 
  \begin{center}
    \begin{tabular}{|c|c|c|c|}
      \hline
Cores segregated? & Fragmentation & Migration? & Figs. \\	
\hline
No & Random & No & \ref{mass_functions},\ref{random_dist}, \ref{lambda_random} \\
Yes & Random &  No & \ref{mass_functions}, \ref{random_dist_MS}, \ref{lambda_MS}, \ref{core_comp} \\
Yes & low-mass & No  & \ref{mass_functions}, \ref{random_dist_MS}, \ref{lambda_MS-a}, \ref{MS_1to5} \\
Yes & high-mass & No & \ref{mass_functions}, \ref{random_dist_MS}, \ref{lambda_MS-a}, \ref{MS_5to1} \\
Yes & Random & Yes & \ref{mass_functions}, \ref{random_dist_MS}, \ref{lambda_MS-a}, \ref{MS_offset}, \ref{core_comp} \\
      \hline
    \end{tabular}
  \end{center}
  \label{sim_summary}
\end{table}

\section{Results}
\label{results}

\subsection{No mass segregation, random fragmentation}

In our first model we randomly distribute the cores within the fractal geometry. We then assume these cores fragment and the fragments from each core remain roughly at the position of each individual core. In our random fragmentation model, each individual core is allowed to fragment into between one and five pieces, and the mass of the fragments is the core mass divided by the number of fragments.

\begin{figure}
 \begin{center}
\rotatebox{270}{\includegraphics[scale=0.35]{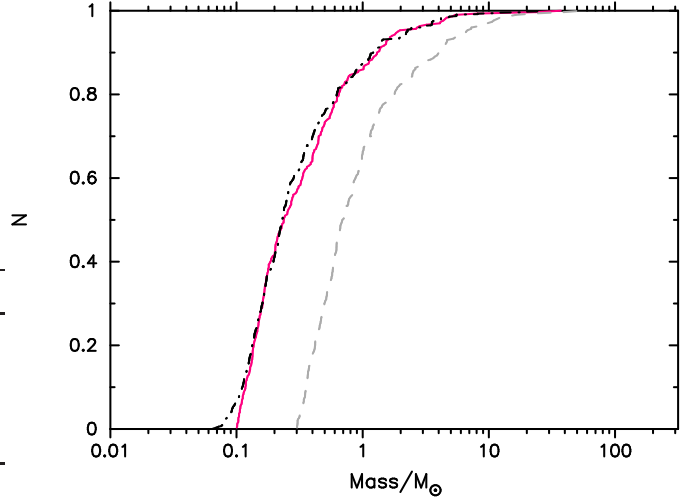}}
\caption[bf]{Cumulative distributions of the mass functions of different objects in our random fragmentation models. The core mass function is shown by the (grey) dashed line; this is the observed stellar IMF (the solid raspberry line) shifted by a factor of three to higher masses. The IMF of the fragments we generate from the core mass function is shown by the dot-dashed black line.}
\label{mass_functions}
\end{center}
 \end{figure}

The mass functions for the distribution of cores and the resultant fragments are shown in Fig.~\ref{mass_functions}. The core mass function, which has the same shape as the \citet{Maschberger13} IMF for stars but is shifted by a factor of three to higher masses, is shown by the grey dashed line. Our random core fragmentation model produces a fragment mass function (FMF) which is shown by the black dot--dashed line. For reference, the \citet{Maschberger13} stellar IMF is shown by the solid raspberry line. 

\begin{figure}
 \begin{center}
\rotatebox{270}{\includegraphics[scale=0.4]{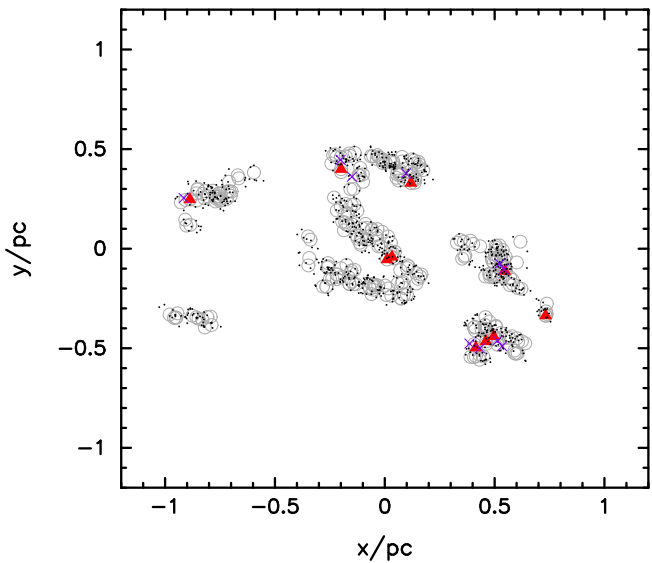}}
\caption[bf]{Distribution of cores and fragments where the the most massive cores are placed randomly within the spatial distribution. The positions of the cores are shown by the grey open circles, and the ten most massive cores are indicated by the solid red triangles. The positions of the fragments are shown by the black points, and the ten most massive fragments are shown by the purple crosses.}
\label{random_dist}
\end{center}
 \end{figure}

In this first model, we randomly distribute the cores in a fractal geometry in an attempt to mimic the filamentary nature of star formation. The spatial distribution of the cores is shown by the grey open circles in Fig.~\ref{random_dist}. The positions of the ten most massive cores are shown by the solid red triangles.

Each core fragments into between one and five pieces, and in this first model they inherit the position of their parent core, but with a small random offset applied to each fragment. The positions of the fragments are shown by the black points in Fig.~\ref{random_dist}, and the ten most massive cores are indicated by the purple crosses. 

Due to the nature of the probability distribution we use to describe the initial core mass function, intermediate- and high-mass cores are rare. If we produce a massive core of say 10\,M$_\odot$, many more cores are sampled from the probability distribution with masses of 1\,M$_\odot$ or less. Therefore, even if we apply a random fragmentation process to each core, unless we split the core into more than 10 fragments, the pieces from each core will still have a significant mass compared to the others in the distribution. For this reason, the locations of the most massive fragments in Fig.~\ref{random_dist} are almost always at the locations of the most massive cores. In Section~\ref{different_fragmentation} we will explore two different mechanisms for fragmenting the cores.

\begin{figure*}
  \begin{center}
\setlength{\subfigcapskip}{10pt}
\hspace*{-1.0cm}\subfigure[$\Lambda_{\rm MSR}$ for the cores]{\label{lambda_random-a}\rotatebox{270}{\includegraphics[scale=0.28]{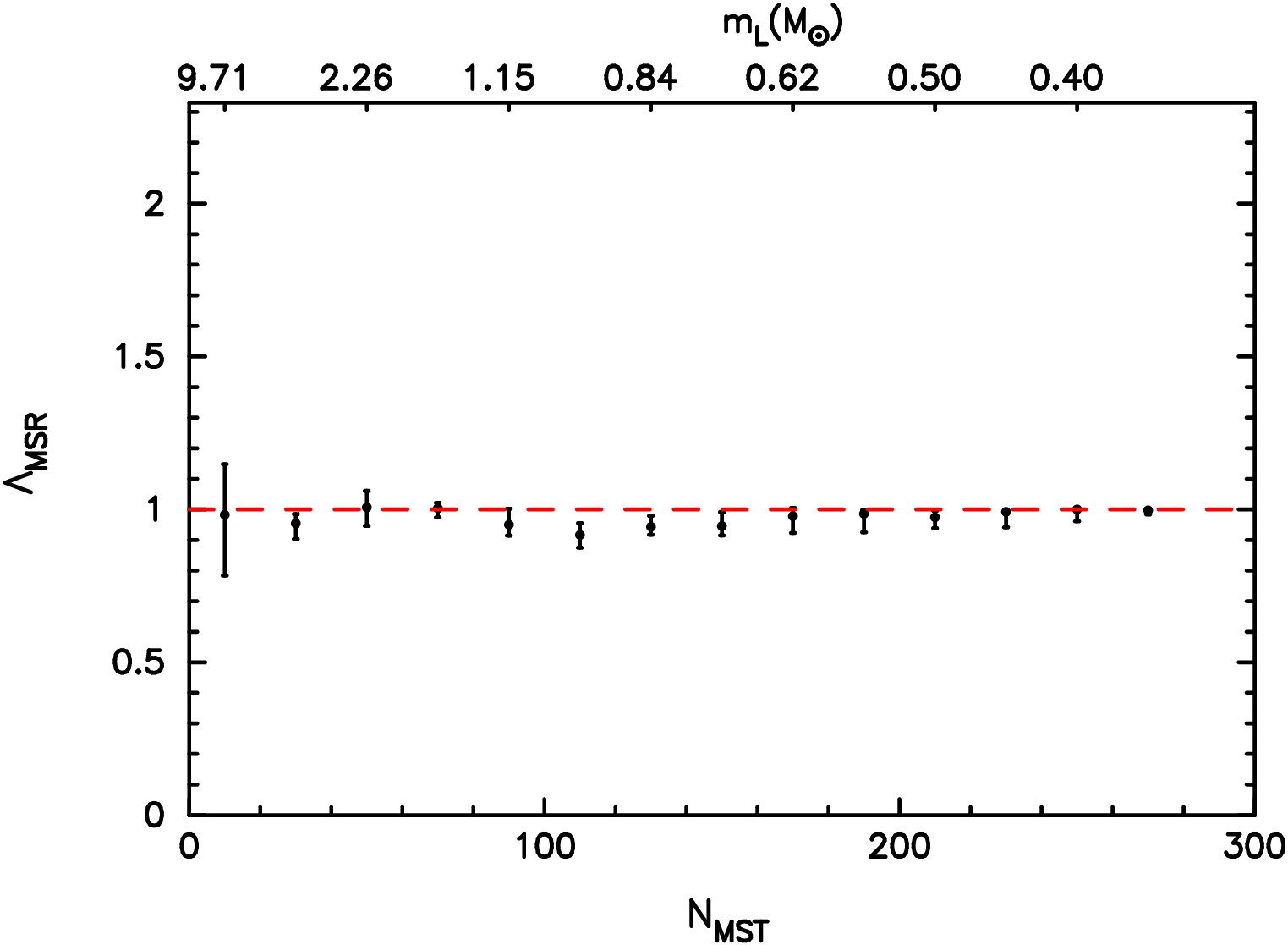}}}
\hspace*{0.3cm}\subfigure[$\Lambda_{\rm MSR}$ for the fragments]{\label{lambda_random-b}\rotatebox{270}{\includegraphics[scale=0.28]{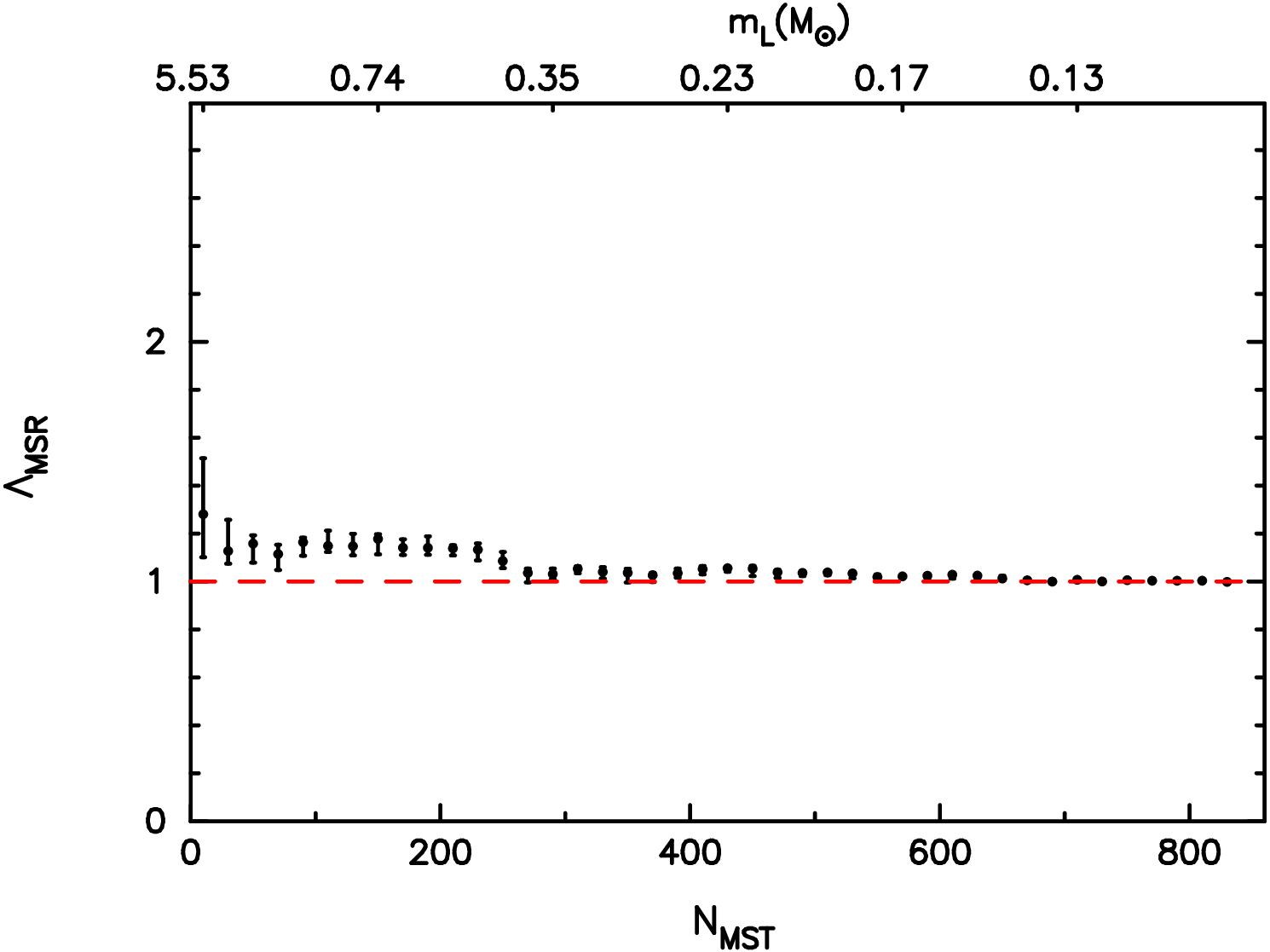}}}  
\caption[bf]{Evolution of the mass segregation ratio $\Lambda_{\rm MSR}$ as a function of the $N_{\rm MST}$ most massive cores (panel a) or fragments (panel b) in the distribution.  The cores are randomly fragmented into between one and five equal-mass pieces. The mass of the least massive object within a set of $N_{\rm MST}$ objects is indicated on the top axis. In this case the most massive cores are distributed randomly within the distribution (the red dashed line indicates $\Lambda_{\rm MSR}= 1$, i.e.\,\,no mass segregation), and the resultant distribution of fragments is also random.}
\label{lambda_random}
  \end{center}
\end{figure*}

We now quantify the relative spatial distributions of the most massive cores and their fragmentation products using the $\Lambda_{\rm MSR} - N_{\rm MST}$ plot in Fig.~\ref{lambda_random}. Panel (a) shows $\Lambda_{\rm MSR}$ as a function of the $N_{\rm MST}$ cores, and panel (b) shows $\Lambda_{\rm MSR}$ as a function of the $N_{\rm MST}$ fragments. The cores are not mass segregated, so $\Lambda_{\rm MSR}$ is consistent with unity. 

Because the most massive cores will often produce the most massive fragments, and these cores are significantly more massive than an average mass core, an individual core often produces two of the most massive fragments in the entire distribution in close proximity to one another. This in turn leads to $\Lambda_{\rm MSR}$ values that are slightly higher than unity (Fig.~\ref{lambda_random-b}). In some realisations of the fractal distribution, these $\Lambda_{\rm MSR}$ values can be significant (i.e.\,\,$\Lambda_{\rm MSR}>2$).

\subsection{Core mass segregation, random fragmentation}

\begin{figure}
 \begin{center}
\rotatebox{270}{\includegraphics[scale=0.4]{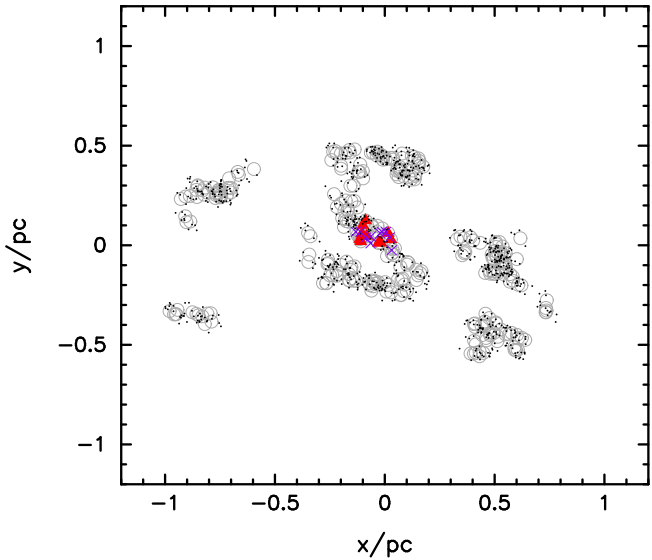}}
\caption[bf]{Distribution of cores and fragments where the the most massive cores are mass segregated. The positions of the cores are shown by the grey open circles, and the ten most massive cores are indicated by the solid red triangles. The positions of the fragments are shown by the black points, and the ten most massive fragments are shown by the purple crosses.}
\label{random_dist_MS}
\end{center}
 \end{figure}

We now mass segregate the cores by ordering them by mass so the most massive core is the most central object, and the least massive core is the least central object. This is shown in Fig.~\ref{random_dist_MS}, where the ten most massive cores (the solid red triangles) are clearly centrally concentrated. 

\begin{figure*}
  \begin{center}
\setlength{\subfigcapskip}{10pt}
\hspace*{-1.0cm}\subfigure[$\Lambda_{\rm MSR}$ for the cores]{\label{lambda_MS-a}\rotatebox{270}{\includegraphics[scale=0.28]{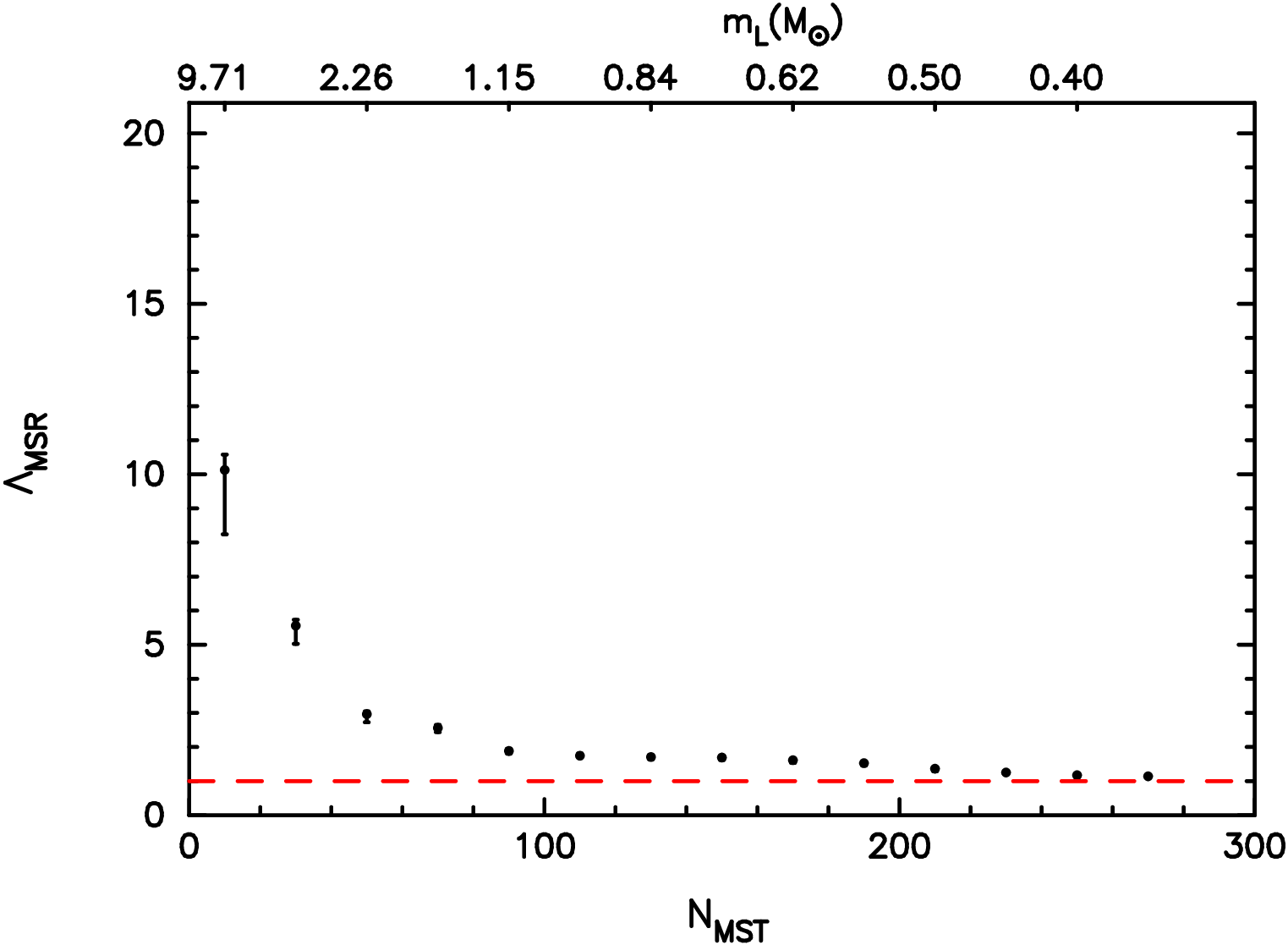}}}
\hspace*{0.3cm}\subfigure[$\Lambda_{\rm MSR}$ for the fragments]{\label{lambda_MS-b}\rotatebox{270}{\includegraphics[scale=0.28]{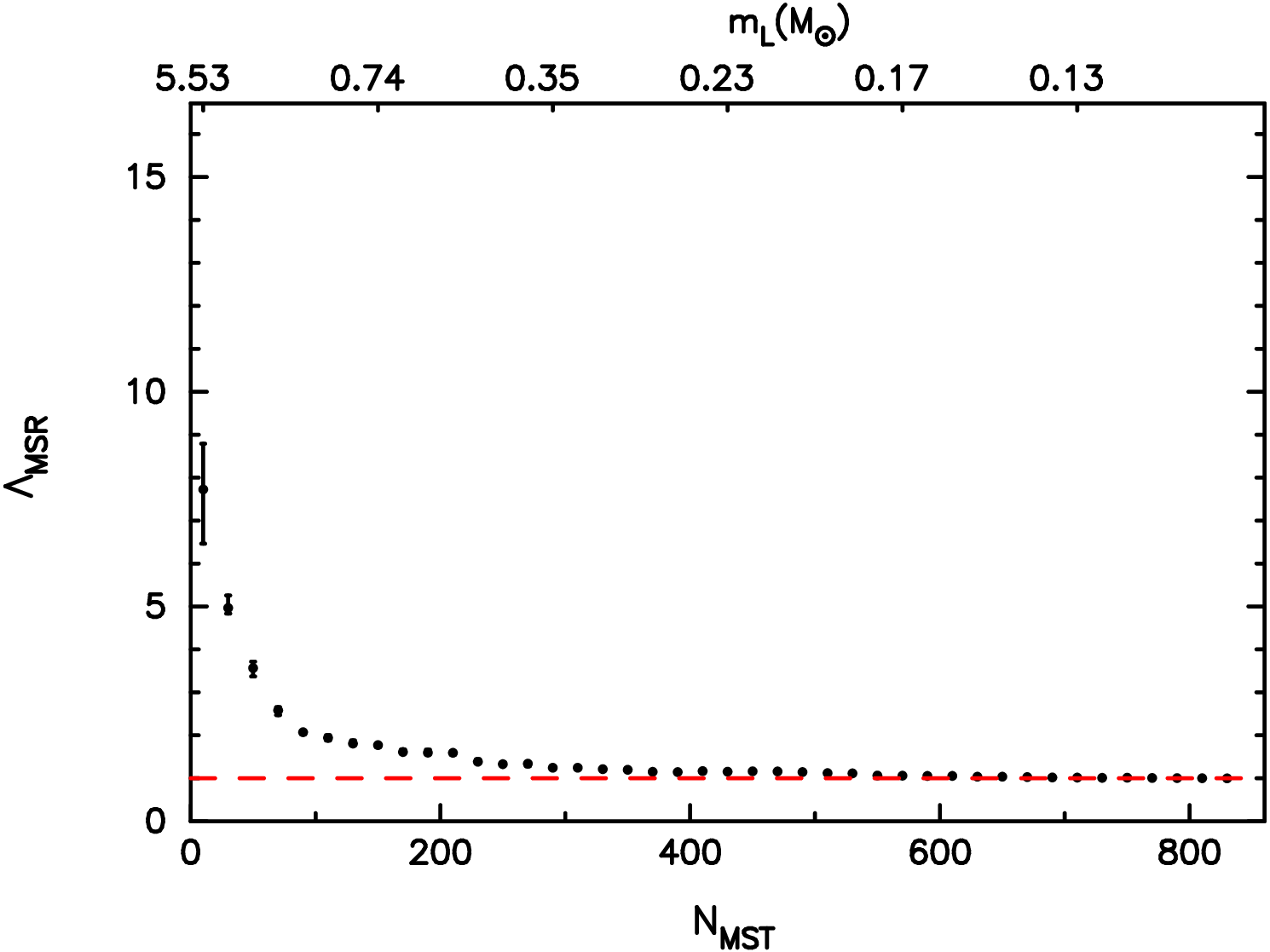}}}  
\caption[bf]{Evolution of the mass segregation ratio $\Lambda_{\rm MSR}$ as a function of the $N_{\rm MST}$ most massive cores (panel a) or fragments (panel b) in the distribution. The cores are randomly fragmented into between one and five equal-mass pieces. The mass of the least massive object within a set of $N_{\rm MST}$ objects is indicated on the top axis. In this case the most massive cores are segregated, and the resultant distribution of fragments is also mass segregated.}
\label{lambda_MS}
  \end{center}
\end{figure*}

The cores are strongly mass segregated, as shown in Fig.~\ref{lambda_MS-a}, where the ten most massive cores have $\Lambda_{\rm MSR} = 10$. The most massive fragments are mostly from the ten most massive cores, and are also very centrally concentrated (the purple crosses in Fig.~\ref{random_dist_MS}). We quantify the level of mass segregation of the fragments in Fig.~\ref{lambda_MS-b}, and although they are not as strongly mass segregated as the cores, the ten most massive fragments still have  $\Lambda_{\rm MSR} = 7.7$.

In all of our distributions in which the cores randomly fragment, the level of mass segregation in the fragments is slightly lower than the level of mass segregation of the cores, but is still significant, and inconsistent with the observations.

\subsubsection{Different ways of fragmenting the cores}
\label{different_fragmentation}

We now trial two slightly different methods for fragmenting the cores. 

\begin{figure*}
  \begin{center}
\setlength{\subfigcapskip}{10pt}
\hspace*{-1.0cm}\subfigure[Mass functions]{\label{lambda_MS_1to5-a}\rotatebox{270}{\includegraphics[scale=0.35]{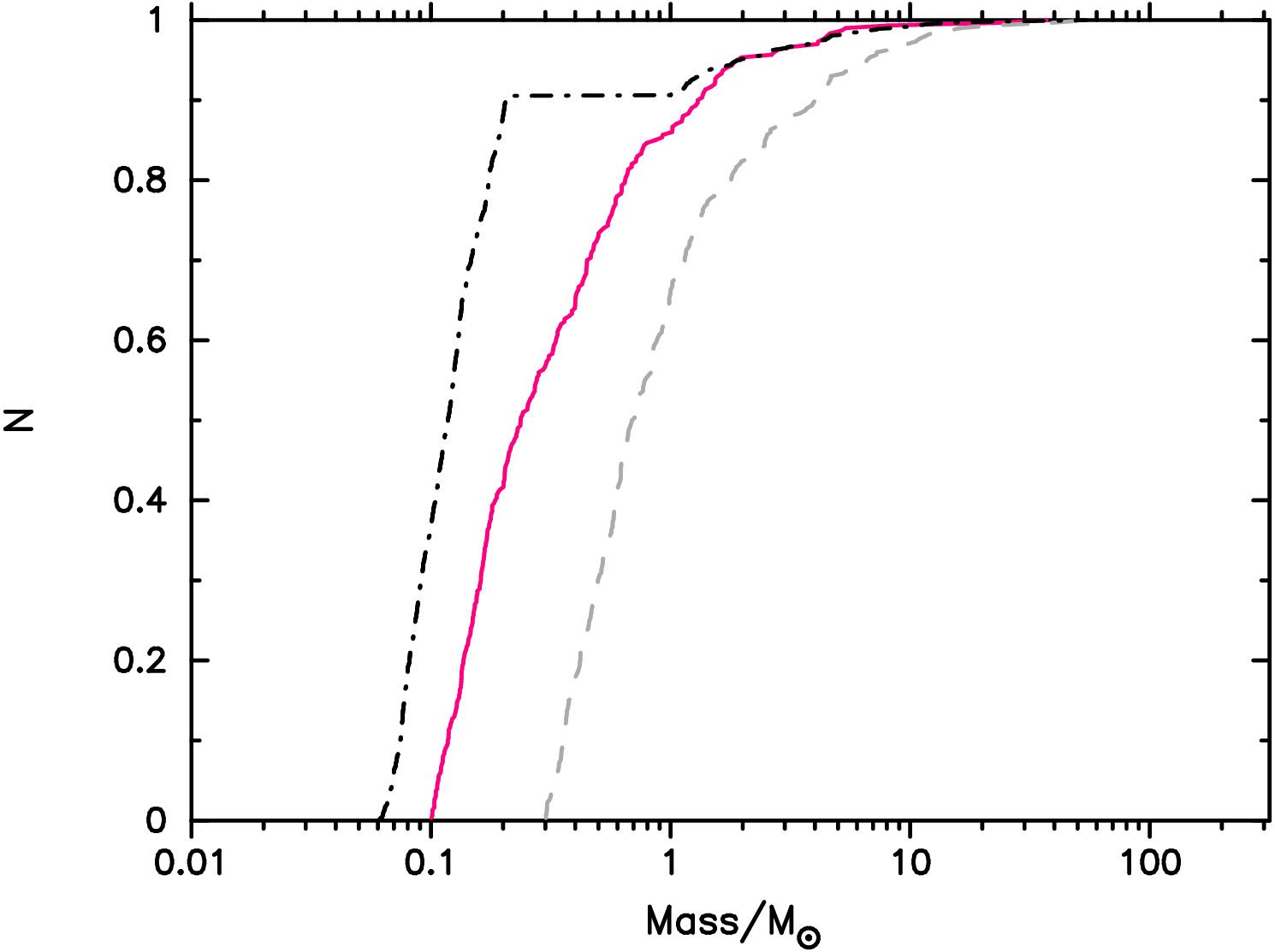}}}
\hspace*{0.3cm}\subfigure[$\Lambda_{\rm MSR}$ for the fragments]{\label{lambda_MS_1to5-b}\rotatebox{270}{\includegraphics[scale=0.33]{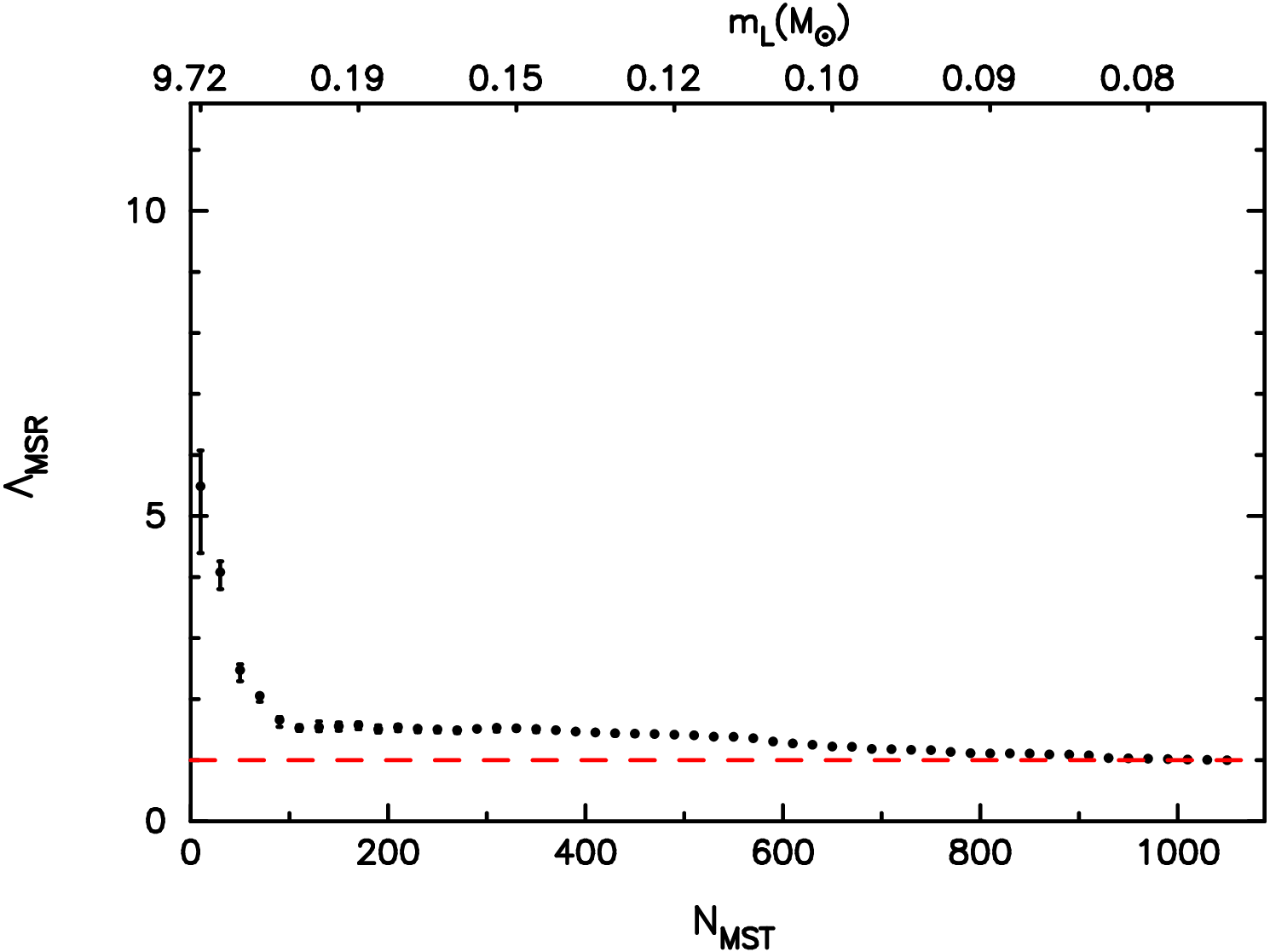}}}  
\caption[bf]{Results for core fragmentation where cores less massive than 1\,M$_\odot$ fragment into five equal mass pieces and cores above this mass do not fragment (i.e.\,\, they form a star with the same mass as the core). Panel (a) shows the mass functions for this scenario; the core mass function is shown by the grey dashed line, and the fragment mass function is shown by the black dot--dashed line. For illustrative purposes, the \citet{Maschberger13} IMF is shown by the solid raspberry line. Panel (b) shows the evolution of the mass segregation ratio $\Lambda_{\rm MSR}$ as a function of the $N_{\rm MST}$ most massive fragments in the distribution (the cores are mass segregated and have exactly the same $\Lambda_{\rm MSR} - N_{\rm MST}$ distribution as in Fig.~\ref{lambda_MS-a}). The mass of the least massive object within a set of $N_{\rm MST}$ objects is indicated on the top axis and $\Lambda_{\rm MSR} = 1$ (no mass segregation) is shown by the red dashed line.}
\label{MS_1to5}
  \end{center}
 \end{figure*}

First, cores that have masses $< 1$M$_\odot$ are fragmented into five equal-mass pieces, whereas more massive cores fragment into only one piece. This is to mimic the effects of any radiation heating that may take place in the protostellar core, which would presumably suppress fragmentation.  The mass function of these fragments in shown in Fig.~\ref{lambda_MS_1to5-a}. Clearly, this artificial method of fragmenting the cores does not reproduce the stellar IMF, but we are interested in whether the over-concentration of low-mass fragments would lead to a neutral mass segregation ratio for the most massive fragments. In Fig.~\ref{lambda_MS_1to5-b} we show the evolution of the $\Lambda_{\rm MSR}$ mass segregation ratio for this bottom-heavy fragmentation, assuming that the cores are mass segregated as in Fig.~\ref{lambda_MS-a}.  Clearly, the most massive fragments still follow the spatial distribution of the cores and this fragmentation does not wipe out mass segregation for the fragments.

\begin{figure*}
  \begin{center}
\setlength{\subfigcapskip}{10pt}
\hspace*{-1.0cm}\subfigure[Mass functions]{\label{lambda_MS_5to1-a}\rotatebox{270}{\includegraphics[scale=0.35]{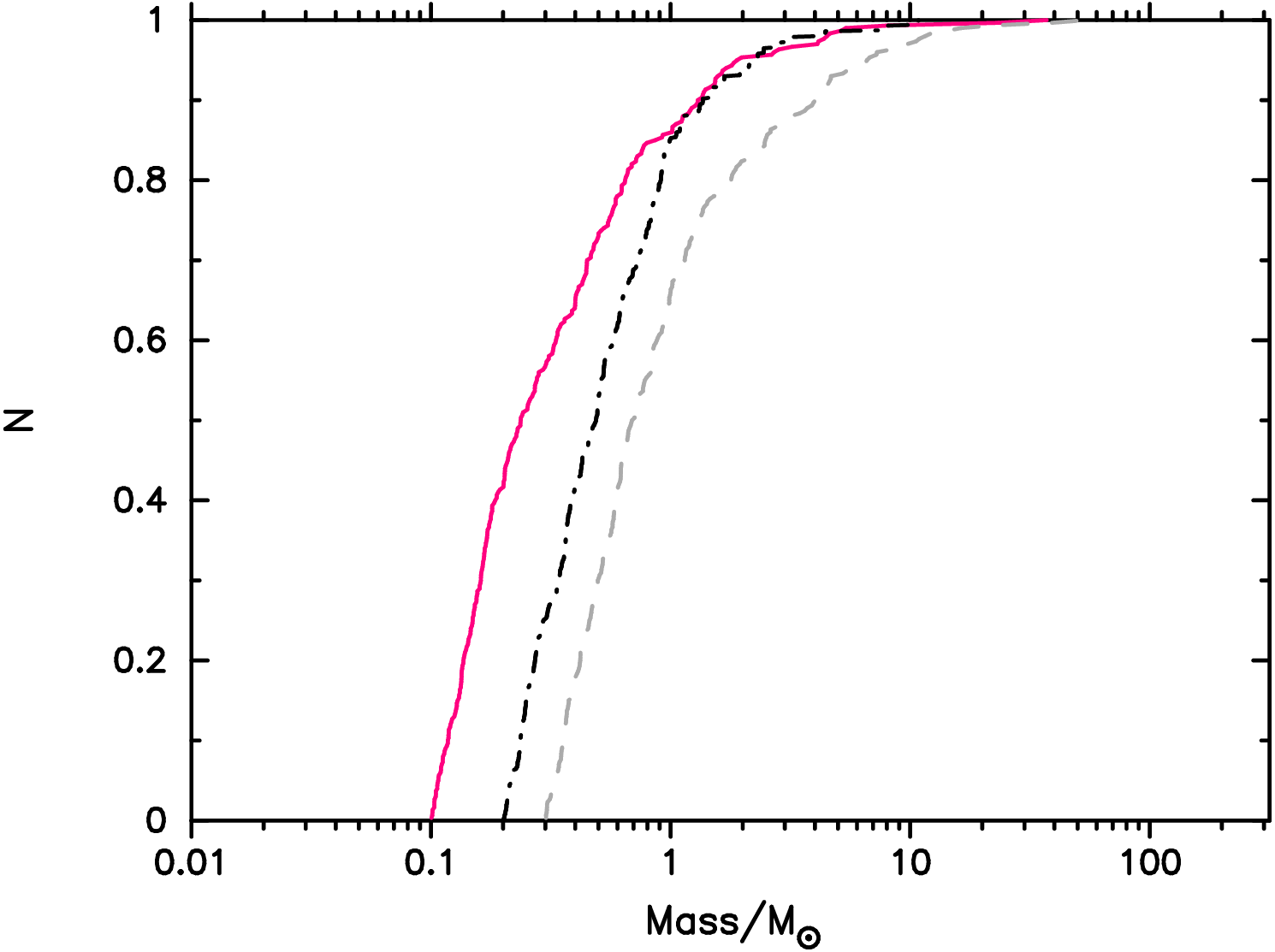}}}
\hspace*{0.3cm}\subfigure[$\Lambda_{\rm MSR}$ for the fragments]{\label{lambda_MS_5to1-b}\rotatebox{270}{\includegraphics[scale=0.33]{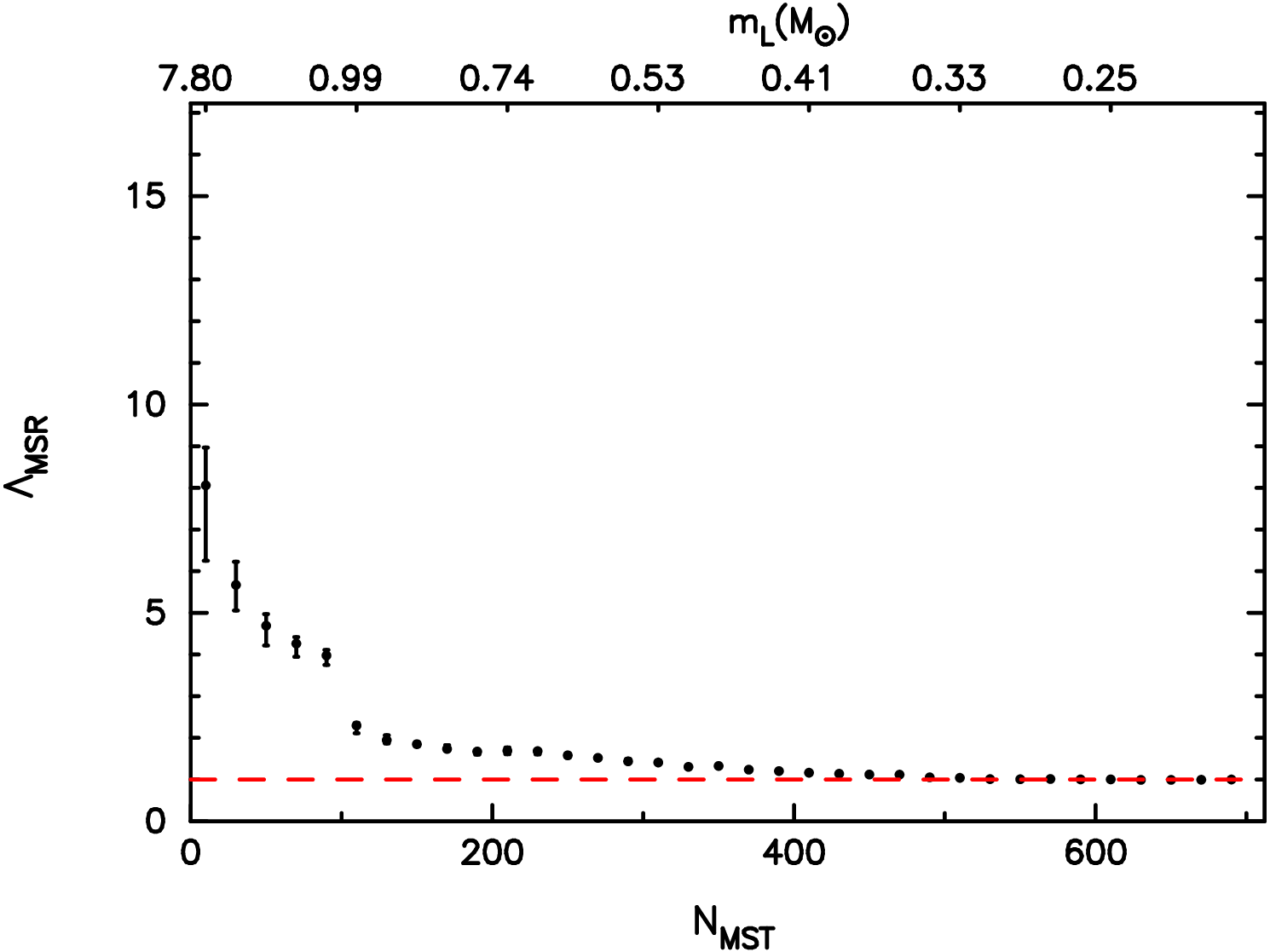}}}  
\caption[bf]{Results for core fragmentation where cores more massive than 1\,M$_\odot$ fragment into five equal mass pieces and cores below this mass do not fragment (i.e.\,\, they form a star with the same mass as the core). This is the inverse of the mechanism shown in Fig.~\ref{MS_1to5}. Panel (a) shows the mass functions for this scenario; the core mass function is shown by the grey dashed line, and the fragment mass function is shown by the black dot--dashed line. For illustrative purposes, the \citet{Maschberger13} IMF is shown by the solid raspberry line. Panel (b) shows the evolution of the mass segregation ratio $\Lambda_{\rm MSR}$ as a function of the $N_{\rm MST}$ most massive fragments in the distribution (the cores are mass segregated and have exactly the same $\Lambda_{\rm MSR} - N_{\rm MST}$ distribution as in Fig.~\ref{lambda_MS-a}). The mass of the least massive object within a set of $N_{\rm MST}$ objects is indicated on the top axis and $\Lambda_{\rm MSR} = 1$ (no mass segregation) is shown by the red dashed line.}
\label{MS_5to1}
  \end{center}
\end{figure*}

 Secondly, we perform the inverse of the above fragmentation mechanism. This time we allow cores $> 1$M$_\odot$ to fragment into five equal-mass pieces, and cores with lower masses form single stars with these masses. In Fig.~\ref{lambda_MS_5to1-a} we show the mass function for these fragments (the black dot--dashed line), which is again inconsistent with the observed stellar IMF. The  $\Lambda_{\rm MSR}$ mass segregation ratio for this top-heavy fragmentation, assuming the cores are mass segregated, is shown in Fig.~\ref{lambda_MS_5to1-b}. Again, the fragments that are produced from the cores are still mass segregated to a significant level.

\subsection{Core mass segregation, random fragmentation followed by dynamical evolution}

Given that any type of core fragmentation increases the amount mass segregation of the stellar fragments with respect to the level of core mass segregation, we next perform a simple thought experiment in which we suppose that the fragments migrate from their formation site due to dynamical evolution.

As before, the initial distribution of cores is mass segregated, and each core is randomly fragmented into between one and five pieces. We then assume some dynamical evolution of the fragments has occurred. We do not directly model the dynamical evolution (e.g.\,\,using $N$-body simulations), but instead apply a random distance offset of order 0.5\,pc to each fragment.

The size of the offset is informed by simulations of the dynamical evolution of pre-main sequence stars in star-forming regions \citep[e.g.][]{Allison10,Parker14b,Sills18}. The typical observed velocity dispersion of cores is low \citep[$\sim$0.3\,km\,s$^{-1}$, e.g.][]{Foster15}. However, the velocity dispersion of pre-main sequence stars in the same regions is higher, at around 1\,km\,s$^{-1}$. This increase in velocity is thought to be due to the dynamical evolution of the pre-main sequence stars in such regions \citep{Parker16b}, and means that young stars can travel pc-scale distances within a Myr (as 1\,km\,s$^{-1}$ roughly translates into 1\,pc\,Myr$^{-1}$). Furthermore, the dynamical decay of hierarchical systems of fragments within cores could also eject pre-main sequence stars considerable distances from their birth sites \citep{Reipurth01,Goodwin07}.

  We assume a random offset to each fragment of 0.5\,pc, and the resultant distribution of the cores and the fragments is shown in Fig.~\ref{lambda_MS_offset-a}. We present results where the random offset is lower (0.25\,pc) in the Appendix.

The cores have the same mass segregation ratio as in Fig.~\ref{lambda_MS-a}. However, because the fragments have been allowed to migrate in random directions, the mass segregation ratio for the fragments is significantly lower (in this example $\Lambda_{\rm MSR,\,fragments}$ is no higher than 2, whereas the core mass segregation ratio peaks at $\Lambda_{\rm MSR,\,cores} = 10$. \\

\begin{figure}
 \begin{center}
\rotatebox{270}{\includegraphics[scale=0.33]{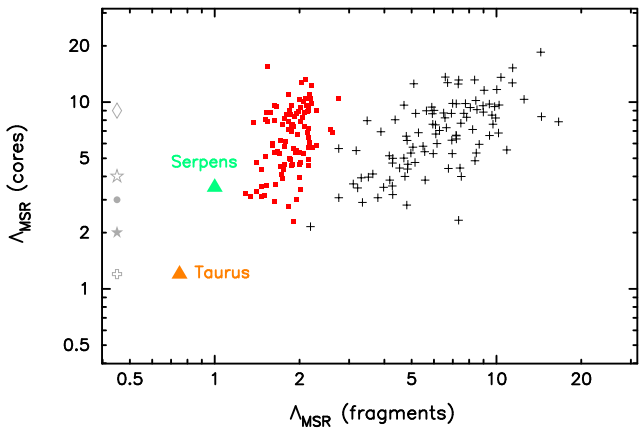}}
\caption[bf]{Comparison of $\Lambda_{\rm MSR}$ for the core distribution against  $\Lambda_{\rm MSR}$ for the fragment distribution. In all cases, the cores are mass segregated. The black plus symbols indicate simulations where the cores have randomly fragmented into between one and five pieces of equal mass. The red squares show simulations where the fragments randomly migrate from their birth sites by up to 0.5\,pc. We show observational data for two regions that have mass segregation measurements for both the cores \emph{and} the stars (Taurus and Serpens). We also show the core segregation measurements in Corona Australis (diamond), NGC2023/2024 (open star), W43 (filled circle), NGC2068/2071 (filled star) and Aquila (open plus), none of which have a corresponding measurement for the pre-main sequence stars. }
\label{core_comp}
\end{center}
 \end{figure}

For our comparison so far, we have only shown the results from one simulation. In Fig.~\ref{core_comp} we show 100 realisations of our simulations where the cores are mass segregated, and then randomly fragment into between one and five pieces (the black plus symbols). We also show 100 realisations of the simulations in which a $\sim$0.5\,pc offset is randomly applied to the fragments (the red squares). In this plot we are showing $\Lambda_{\rm MSR}$ for the ten most massive cores versus $\Lambda_{\rm MSR}$ for the ten most massive fragments.

Clearly, if the cores are already mass segregated, the resultant distribution of fragments will also be mass segregated, unless the fragments are allowed to migrate within the star-forming region.

In Fig.~\ref{core_comp} we also show the observational data for two star-forming regions for which $\Lambda_{\rm MSR}$ has been calculated for both the cores and pre-main sequence stars. The data for Serpens come from \citet{Plunkett18} and the data for Taurus come from \citet{Parker11b} for the pre-main sequence stars and \citet{Dib18} for the cores. We also show the $\Lambda_{\rm MSR}$ values for cores in regions where there are no available measurements for the pre-main sequence stars; Corona Australis \citep[][the open diamond]{Dib18}, NGC2023/2024 in Orion B \citep[][the open star]{Parker18a}, W43 \citep[][the filled circle]{Dib18}, NGC2068/2071 in Orion B \citep[][the filled star]{Parker18a} and Aquila \citep[][the open plus]{Dib18}.

  Although not all the observed regions show mass segregation of their cores, in both regions where we measure mass segregation for the pre-main sequence stars (Serpens and Taurus), the amount of mass segregation is lower for the pre-main sequence stars than for the cores \citep[in Taurus, the pre-main sequence stars are actually \emph{inversely} mass-segregated,][]{Parker11b}.

Furthermore, the amount of initial core mass segregation in our toy models is consistent with the amount of mass segregation in the cores in five of the seven regions for which $\Lambda_{\rm MSR, cores}$ has been measured. However, measurements of $\Lambda_{\rm MSR}$ for the pre-main sequence stars are required for every region in order to fully test our model.

 \begin{figure*}
  \begin{center}
\setlength{\subfigcapskip}{10pt}
\hspace*{-1.0cm}\subfigure[Distribution of cores and dispersed fragments]{\label{lambda_MS_offset-a}\rotatebox{270}{\includegraphics[scale=0.35]{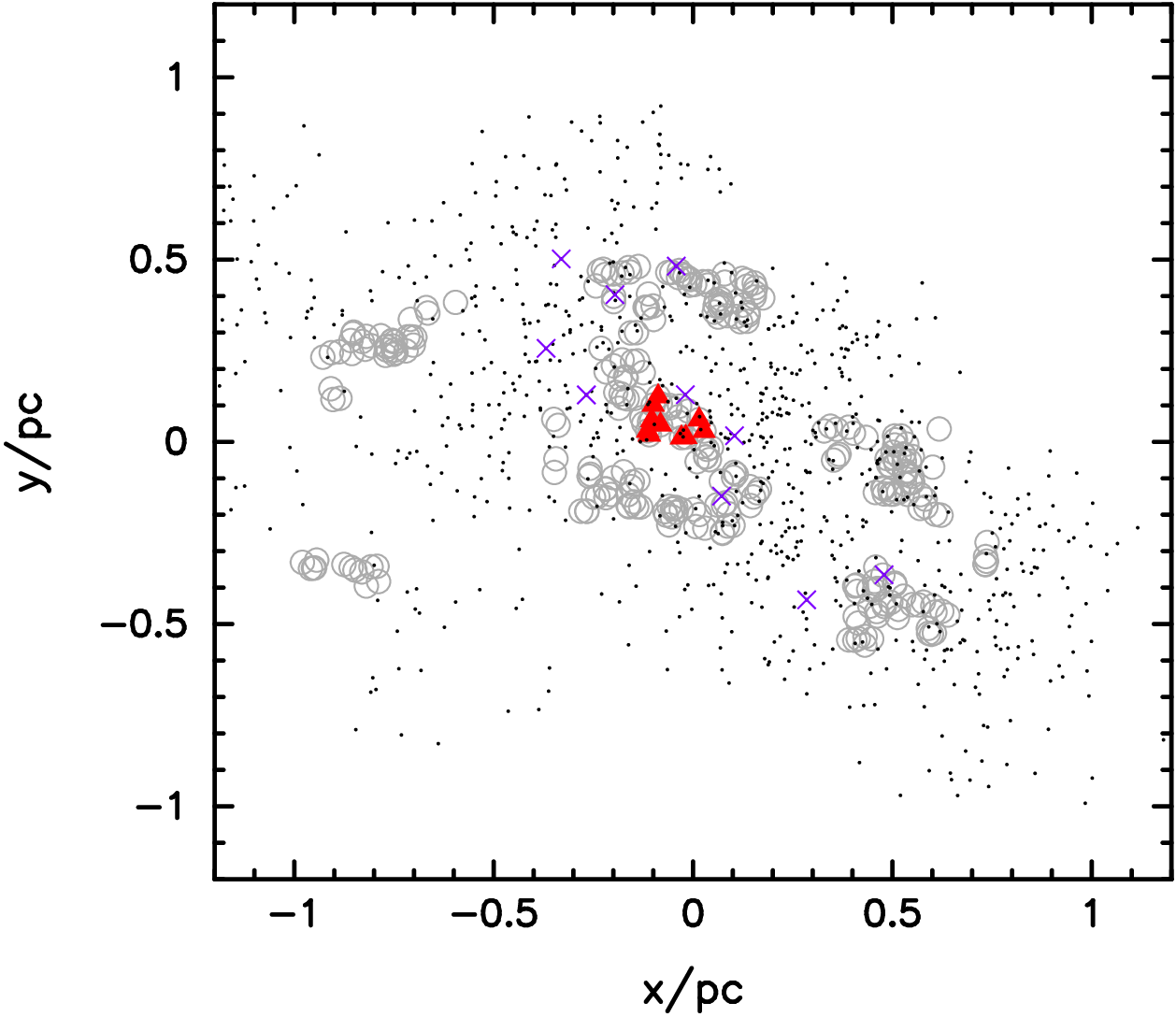}}}
\hspace*{0.3cm}\subfigure[$\Lambda_{\rm MSR}$ for the dispersed fragments]{\label{lambda_MS_offset-b}\rotatebox{270}{\includegraphics[scale=0.33]{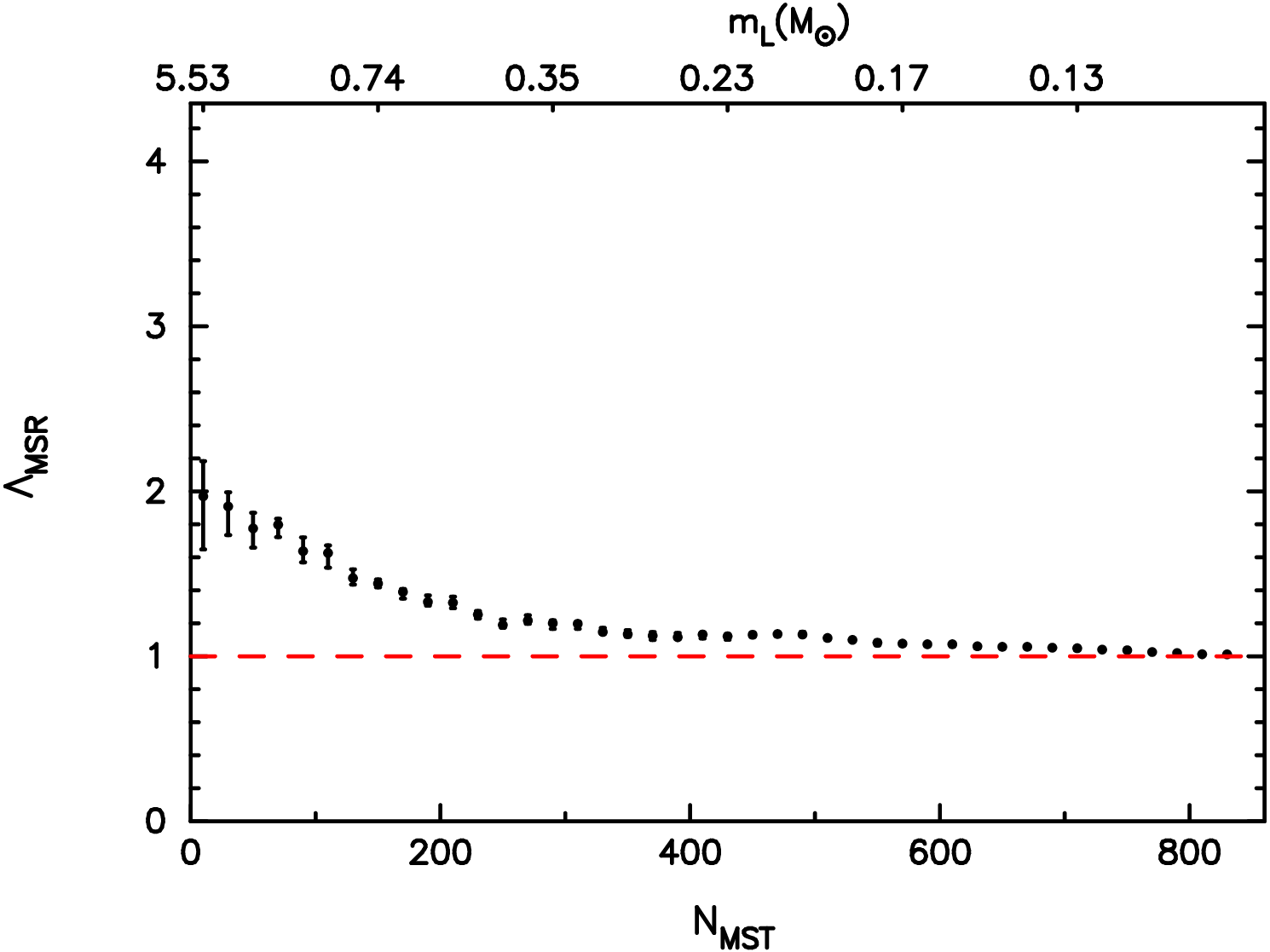}}}  
\caption[bf]{The positions of mass segregated cores and the fragments produced by these cores after applying a spatial offset of up to 0.5\,pc to the fragments to mimic the effects of dynamical evolution (panel a).  The positions of the cores are shown by the grey open circles, and the ten most massive cores are indicated by the solid red triangles. The positions of the fragments are shown by the black points, and the ten most massive fragments are shown by the purple crosses. Panel (b) shows the evolution of the mass segregation ratio $\Lambda_{\rm MSR}$ as a function of the $N_{\rm MST}$ most massive fragments in the distribution (the cores are mass segregated and have exactly the same $\Lambda_{\rm MSR} - N_{\rm MST}$ distribution as in Fig.~\ref{lambda_MS-a}). The mass of the least massive object within a set of $N_{\rm MST}$ objects is indicated on the top axis.}
\label{MS_offset}
  \end{center}
\end{figure*}

     \section{Discussion}
\label{discuss}


The random fragmentation algorithm we use to convert the cores into stars remarkably reproduces the stellar initial mass function (Fig~\ref{mass_functions}). However, the properties of many multiple systems almost certainly derive from the fragmentation process in cores \citep{Goodwin07,Goodwin09b,Holman13,Lomax14}. Our random fragmentation algorithm splits cores into equal-mass fragments, implying that if these fragments subsequently remained as bound multiple systems, their mass ratio distributions would peak at unity. The mass ratio distribution for binary systems in the field is flat \citep{Reggiani11a,Reggiani13}, whereas the mass ratios of components of triples and quadruples tend to be closer to unity \citep{Tokovinin08}. Nevertheless, this warrants further investigation and in a future paper, we will consider the fragmentation process of cores with the goal of reproducing the observed multiplicity properties in star-forming regions and the Galactic field. 

Our simulations clearly show that the fragmentation of mass segregated cores cannot produce a non-mass segregated distribution of fragments.  The main reason for this is that even if a fairly massive core (e.g.\,\,10\,M$_\odot$) fragments into 5 stars, those stars are still relatively `high' mass ($\sim 2$\,M$_\odot$) compared to the average stellar mass. Furthermore, if those fragments do not migrate a large distance from the core site, they automatically assume a very centrally concentrated distribution, which would be observed as a strong mass segregation signature.


One possible caveat is that the method used to detect cores may be biased towards finding cores in a mass segregated distribution. For this bias to influence the observations of \citet{Kirk16b,Alfaro18,Dib18b,Parker18a,Plunkett18}, low-mass cores would have to evade detection in the regions where the most massive cores are readily found. As most of these studies used the $\Lambda_{\rm MSR}$ technique, which assumes no underlying spatial geometry, it is difficult to envisage  why low-mass cores would not be detected in the areas where high-mass cores are located.

We also cannot rule out the possibility that the cores that are observed today follow a different spatial distribution to the cores that formed the pre-main sequence stars. It is possible that feedback from the more massive stars in a region could disrupt the distribution of gas \citep[e.g.][]{Dale12a}, or even photoevaporate nearby cores \citep{Whitworth04}. However, many of the nearby star-forming regions where core mass segregation is observed contain few stars massive enough to ionise their surroundings. Nevertheless, for regions such as Orion \citep{Kirk16b,Parker18a}, this remains a possibility.

In our simulations where we allowed the fragments to migrate from their birth sites, the mass segregation signature of the cores is not observed in the distribution of fragments. $N$-body simulations show that for modest stellar densities ($10 - 100$\,M$_\odot$\,pc$^{-3}$) stars migrate significant distances within Myr timescales.  We therefore suggest that an explanation of the dearth of mass segregation in pre-main sequence stars in star-forming regions where the cores \emph{are} mass segregated may be due to dynamical evolution.

A potential failure of the above migration hypothesis would be if the pre-main sequence stars were found in gas-rich areas of the star-forming region, indicating that they hadn't moved far from their birth environs. However, the process of star-formation tends to use up much of the available gas \citep[so-called `gas exhaustion',][and references within]{Kruijssen12a,Longmore14} and so the absence of gas would not necessarily mean that the pre-main sequence stars had moved significant distances after formation.

\section{Conclusions}
\label{conclude}

We have presented a simple model for the fragmentation of pre-/proto-stellar cores in star-forming regions in an attempt to explain why the cores are often mass segregated, whereas the pre-main sequence (PMS) stars in the same regions typically are not mass segregated. Our conclusions are as follows:

(i) If the distribution of cores is mass segregated, any reasonable approximation for the fragmentation leads to a similar or  higher degree of mass segregation in the distribution of fragments. This is due to the nature of the core mass function; if a massive core (e.g.\,\,10\,M$_\odot$) fragments into five equal mass pieces, this results in an over-density of relatively massive objects, which in turn enhances the mass segregation signal of the fragments.

(ii) A possible explanation for the lack of mass segregation in pre-main sequence stars in star-forming regions where the cores are mass segregated is that the PMS stars have moved signficant distance from their birth sites, and any primordial mass segregation has been erased.

(iii) The biggest caveat in this scenario is that the spatial distribution of cores observed in a given star-forming region may be different to the spatial distribution of cores that formed the PMS population of stars. This could be due to feedback from the most massive stars, and could be tested by analysing star formation in hydrodynamical simulations which include feedback mechanisms. 

\section*{Acknowledgements}

Our thanks to the anonymous referee for their helpful comments and suggestions, which improved the original manuscript. RJP acknowledges support from the Royal Society in the form of a Dorothy Hodgkin Fellowship.

\bibliographystyle{mnras}  
\bibliography{general_ref}

\appendix

\section{Reducing the distance travelled by fragments due to dynamical evolution}
\label{appendix}

In Fig.~\ref{MS_offset} we apply a 0.5\,pc offset to the fragments to mimic the effects of dynamical evolution. The typical velocity dispersion in a star-forming region is $\sim$1\,km\,s$^{-1}$, which is roughly 1\,pc\,Myr$^{-1}$. Whilst the velocity dispersion can be much lower for pre-stellar cores \citep{Foster15}, pre-main sequence stars typically migrate through the star-forming region and can cover distances of up to 1\,pc within several Myr \citep{Parker14b}.

  Nevertheless, the assumption that the pre-main sequence stars have moved by up to 0.5\,pc from their birth locations does impact our results, and here we repeat the analysis presented in Figs.~\ref{MS_offset}~and~\ref{core_comp} but assume a smaller offset -- up to 0.25\,pc instead of up to 0.5\,pc.

\begin{figure*}
  \begin{center}
\setlength{\subfigcapskip}{10pt}
\hspace*{-1.0cm}\subfigure[Distribution of cores and dispersed fragments]{\label{appendix:lambda_MS_offset-a}\rotatebox{270}{\includegraphics[scale=0.35]{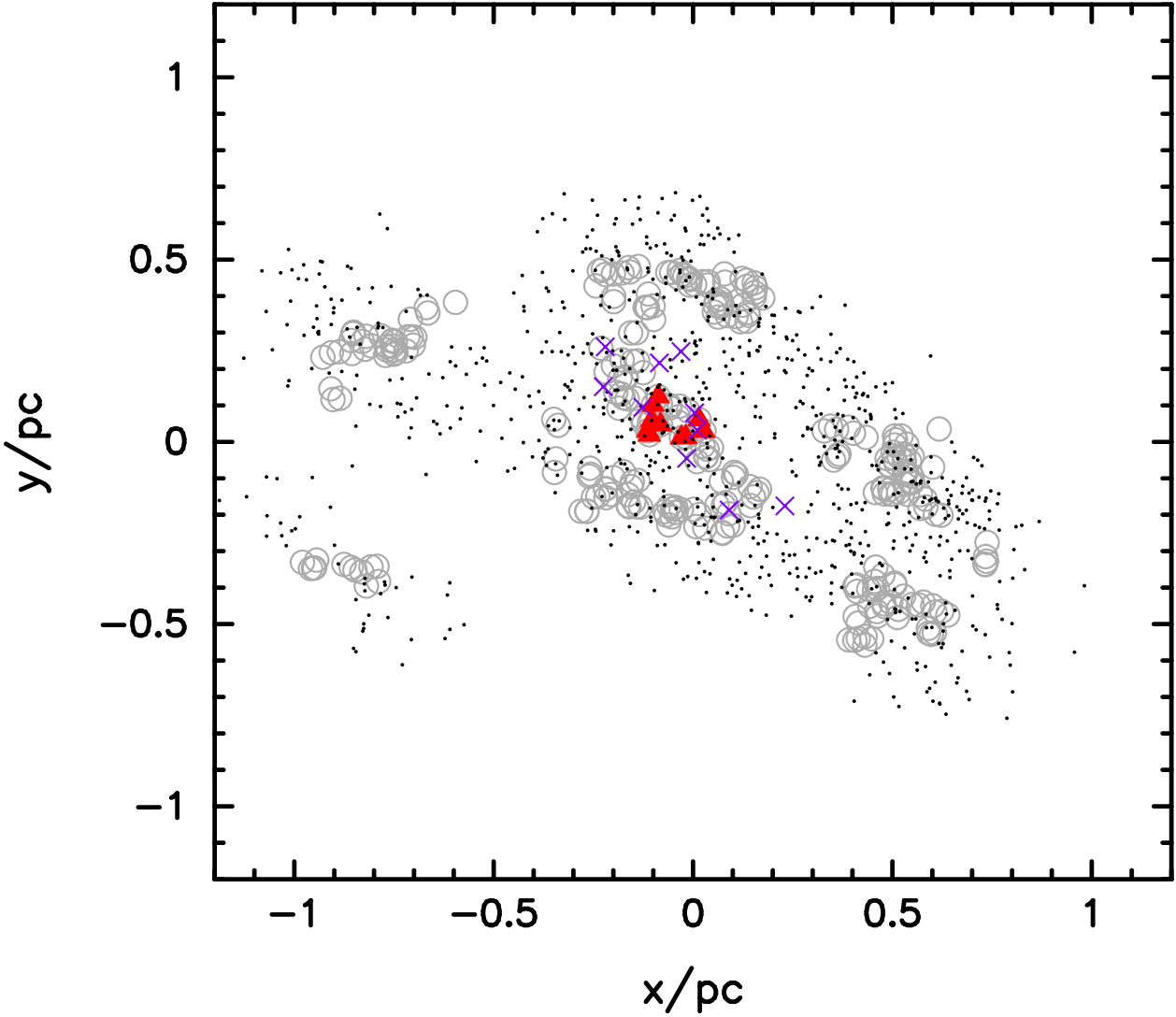}}}
\hspace*{0.3cm}\subfigure[$\Lambda_{\rm MSR}$ for the dispersed fragments]{\label{appendix:lambda_MS_offset-b}\rotatebox{270}{\includegraphics[scale=0.33]{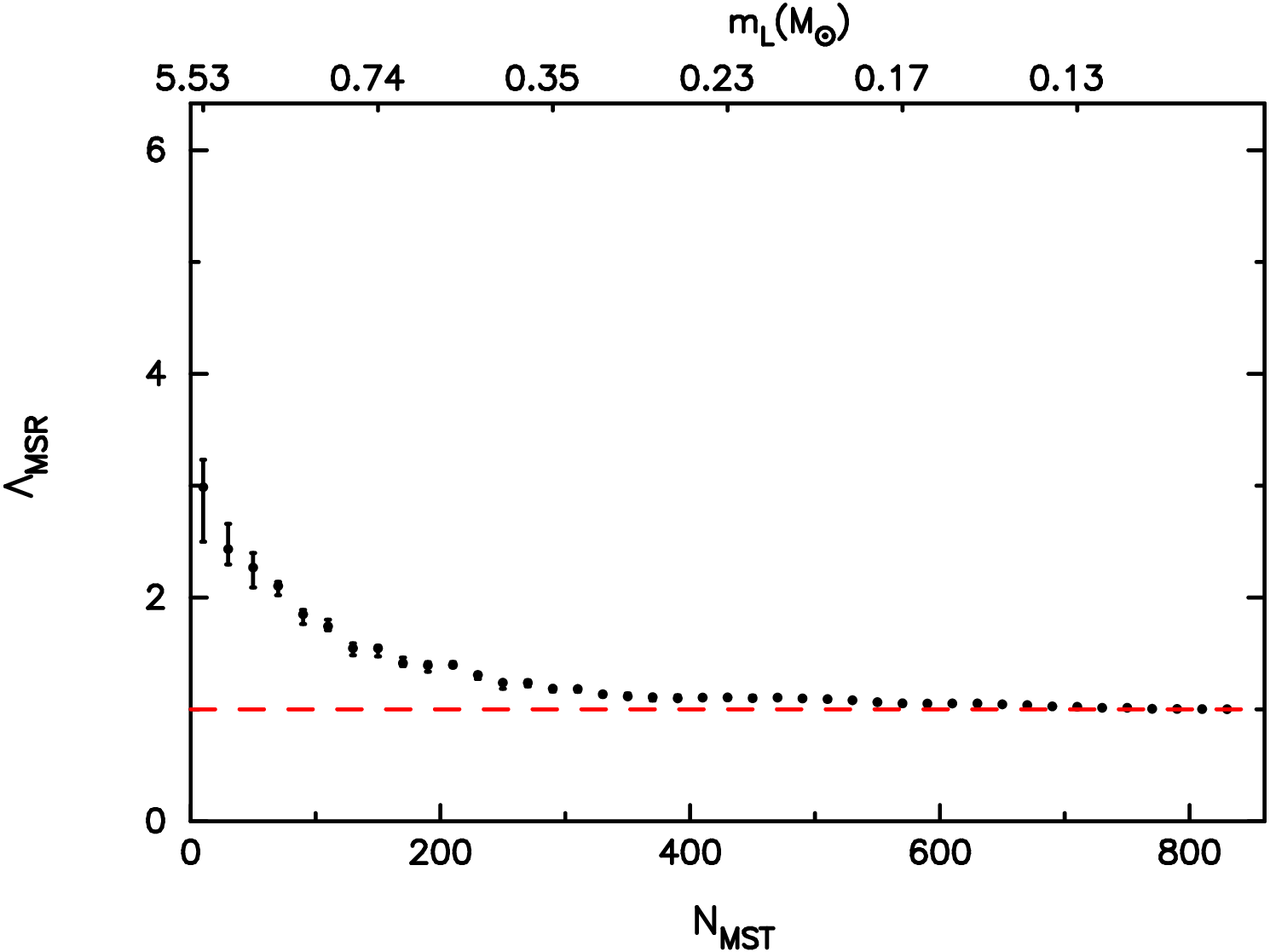}}}  
\caption[bf]{The positions of mass segregated cores and the fragments produced by these cores after applying an spatial offset of up to 0.25\,pc to the fragments to mimic the effects of dynamical evolution (panel a).  The positions of the cores are shown by the grey open circles, and the ten most massive cores are indicated by the solid red triangles. The positions of the fragments are shown by the black points, and the ten most massive fragments are shown by the purple crosses. Panel (b) shows the evolution of the mass segregation ratio $\Lambda_{\rm MSR}$ as a function of the $N_{\rm MST}$ most massive fragments in the distribution (the cores are mass segregated and have exactly the same $\Lambda_{\rm MSR} - N_{\rm MST}$ distribution as in Fig.~\ref{lambda_MS-a}). The mass of the least massive object within a set of $N_{\rm MST}$ objects is indicated on the top axis.}
\label{appendix:MS_offset}
  \end{center}
\end{figure*}

\begin{figure}
 \begin{center}
\rotatebox{270}{\includegraphics[scale=0.33]{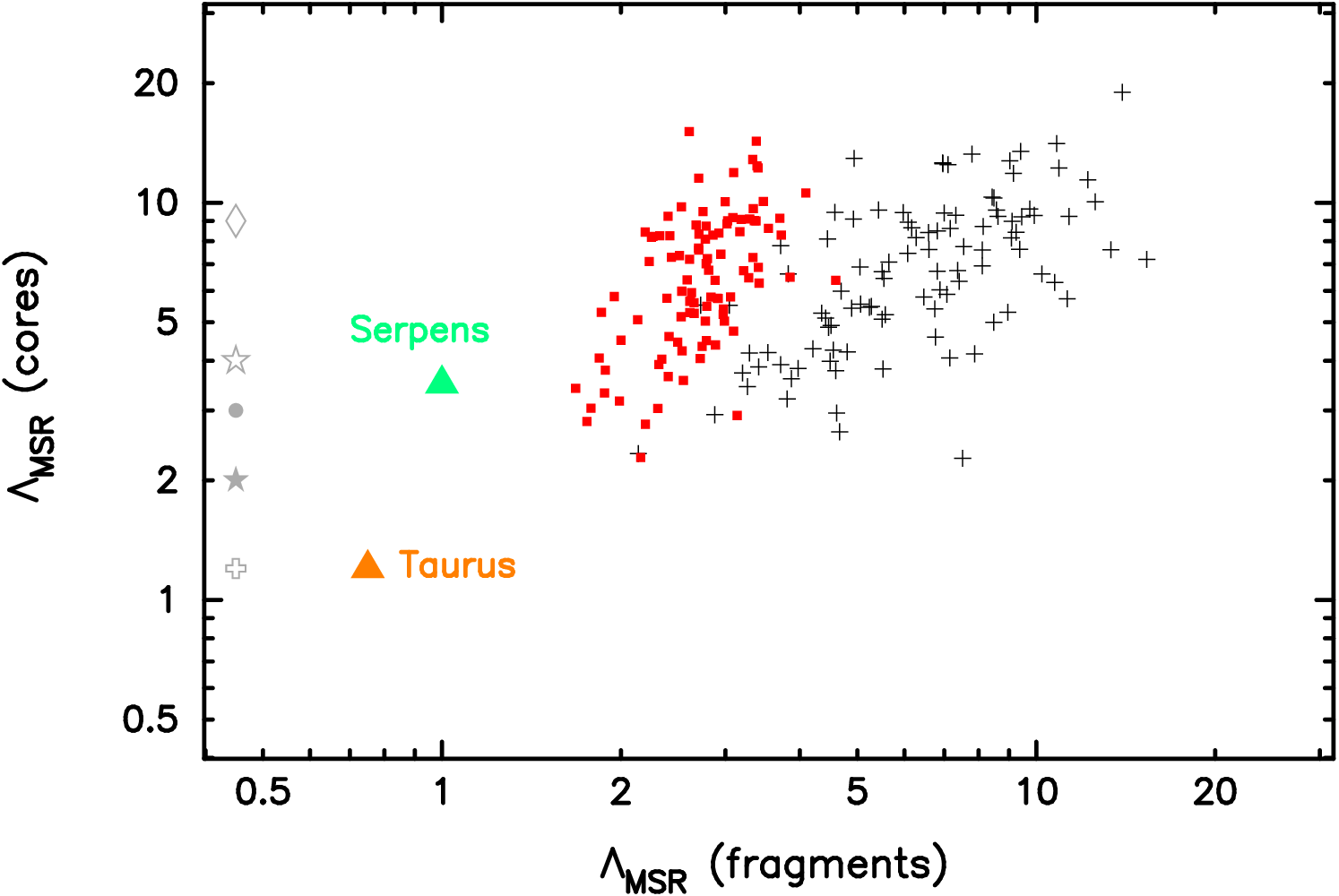}}
\caption[bf]{Comparison of $\Lambda_{\rm MSR}$ for the core distribution against  $\Lambda_{\rm MSR}$ for the fragment distribution. In all cases, the cores are mass segregated. The black plus symbols indicate simulations where the cores have randomly fragmented into between one and five pieces of equal mass. The red squares show simulations where the fragments randomly migrate from their birth sites by up to 0.25\,pc. We show observational data for two regions that have mass segregation measurements for both the cores \emph{and} the stars (Taurus and Serpens). We also show the core segregation measurements in Corona Australis (diamond), NGC2023/2024 (open star), W43 (filled circle), NGC2068/2071 (filled star) and Aquila (open plus), none of which have a corresponding measurement for the pre-main sequence stars. }
\label{appendix:core_comp}
\end{center}
 \end{figure}

  The evolution of $\Lambda_{\rm MSR, fragments}$ is shown in Fig.~\ref{appendix:MS_offset}. Whilst the amount of mass segregation of the fragments is higher than when we assume a larger offset, the level of mass segregation is still much lower than for the cores (compare this with $\Lambda_{\rm MSR, cores}$ in Fig.~\ref{lambda_MS-a}). This is also demonstrated in Fig.~\ref{appendix:core_comp}, where the spatial offset due to dynamical evolution still significantly reduces the mass segregation of the fragments with respect to the cores (compare this with Fig.~\ref{core_comp}).

\label{lastpage}

\end{document}